\documentclass[letterpaper,12pt,onecolumn]{IEEEtran}

\IEEEoverridecommandlockouts

\usepackage{graphicx}
\usepackage{psfrag}

\usepackage[cmex10]{amsmath}
\usepackage{amssymb}
\usepackage{amsthm}
\usepackage{amsfonts}

\usepackage{cases}
\usepackage{units}
\usepackage{dsfont}

\usepackage{cite}
\usepackage{algorithm}
\usepackage{algpseudocode}

\newcommand{\E}{\mathrm{E}}
\newcommand{\st}{\,|\,}

\DeclareMathOperator{\dB}{dB}
\DeclareMathOperator{\tr}{\mathbf{tr}}
\DeclareMathOperator{\diag}{\mathbf{diag}}

\providecommand{\abs}[1]{\lvert{#1}\rvert}

\newtheorem{prop}{Proposition}

\title{Linear Precoding in Cooperative MIMO Cellular Networks with Limited Coordination Clusters}

\author{Chris~T.~K.~Ng and Howard~Huang% stops space
\thanks{The material in this paper was presented in part
at the Forty-Seventh Annual Allerton Conference on Communication, Control, and Computing, in Monticello, IL, September/October, 2009.}%
\thanks{
C.~Ng and H.~Huang are with Bell Laboratories, Alcatel-Lucent, Holmdel, NJ 07733 USA (e-mail: Chris.Ng@alcatel-lucent.com; Howard.Huang@alcatel-lucent.com).}%
}

\begin{document}

\maketitle
\thispagestyle{empty}

%%% ============================================================
\begin{abstract}

In a cooperative multiple-antenna downlink cellular network, maximization of a concave function of user rates is considered.
A new linear precoding technique called soft interference nulling (SIN) is proposed, which performs at least as well as zero-forcing (ZF) beamforming.
All base stations share channel state information,
but each user's message is only routed to those that participate in the user's coordination cluster.
SIN precoding is particularly useful when clusters of limited sizes overlap in the network, in which case traditional techniques such as dirty paper coding or ZF do not directly apply.
The SIN precoder is computed by solving a sequence of convex optimization problems.
SIN under partial network coordination can outperform ZF under full network coordination at moderate SNRs.
Under overlapping coordination clusters,
SIN precoding achieves considerably higher throughput compared to myopic ZF, especially when the clusters are large.

\end{abstract}


%%% ============================================================
\section{Introduction}
\label{sec:intro}

Interference management is a fundamental challenge in wireless cellular systems.
In this paper, we consider the downlink cellular network, and investigate the performance benefits of allowing cooperation and joint processing among the base stations.
Without base station cooperation, the system is interference-limited, i.e., the signal-to-interference-plus-noise ratio (SINR) at the mobiles cannot be improved simply by increasing the base station transmit power, since higher transmit power also creates stronger interference.
Given the deployment of a fixed number of base stations and antennas, one approach to increase system throughput is to allow the joint encoding of user signals across the base stations.
In this case, assuming perfect cooperation among the base stations, the downlink system can be modeled as a broadcast channel (BC).
However, the theoretically optimal dirty paper coding (DPC) transmission scheme for the BC can be too complex for practical implementation.
Zero-forcing (ZF) beamforming is a simple linear precoding technique that offers good performance in a BC\@.
In this paper, we propose a new linear precoding technique called soft interference nulling (SIN) that performs better than or equal to ZF\@.
The SIN precoder can be found by solving a sequence convex optimization problems.
The SIN precoding technique applies to the case when the terminals have multiple antennas,
as well as the case
when the users are served by overlapping coordination clusters in the network, where each coordination cluster consists of a limited number of cooperating base stations.
SIN precoding is particularly useful in the setting of overlapping clusters, since under this scenario traditional precoding techniques such as DPC or ZF cannot be applied directly.

In wireless communications, multiple-input multiple-output (MIMO) transmission techniques
have been shown to provide substantial improvement in channel capacity \cite{foschini98:wcomm_mult_ant, telatar99:cap_mimo_gaus}.
For cellular downlink networks,
the system throughput using time division multiplexing access (TDMA), ZF, and DPC are compared in \cite{viswanathan03:dl_cap_cell_itfr, sharif07:ts_dpc_bf_mimo_bc_users}, and the performance of ZF is studied in \cite{sharif05:mimo_bc_part_si, yoo06:opt_ma_bc_zf_bf}.
Moreover, ZF is generalized for multi-user MIMO channels in \cite{choi04:pre_mu_mimo_decomp, spencer04:zf_dl_mu_mimo}.
Different precoding schemes for MIMO BCs are presented in \cite{caire03:ach_guas_mimo_bc, peel05:vec_pert_mimo_inv_reg}.
The optimality of DPC in a MIMO BC is shown in \cite{weingarten06:cap_mimo_bc}.
For single-cell multiuser MIMO channels, the optimization of different performance metrics in terms of the user rates or SINRs are considered in
\cite{kobayashi06:iter_wf_wtsr_mimo_bc, stojnic06:rate_max_bc_lin_proc, christensen09:wt_srate_mmse_mimo_bc, wiesel06:lin_pre_conic_opt_mimo, shi08:rate_opt_mu_mimo_lin_proc}.
In \cite{sadek07:leakage_precod_dl_mu_mimo}, the transmit precoder is designed to maximize signal strength relative to the interference it causes.
Cooperating base stations with overlapping coordination clusters in the cellular uplink channel is considered in \cite{venkatesan07:coop_uplink_spectral}.
When the user signals are jointly encoded by separate base stations, they are under per-antenna power constraints (PAPC).
ZF under PAPC are considered in \cite{boccardi06:zf_mimo_bc_papc, karakayali07:opt_zf_papc, wiesel08:zf_pre_gen_inv},
and DPC under PAPC is treated in \cite{yu07:tx_opt_ma_dl_papc}.
System-level performance gains in collaborative networks under full-network coordination are studied in \cite{karakayali06:net_coord_cell, venkatesan09:wimax_net_mimo_indoor}.

We consider the maximization of a general concave utility function of user rates in a cellular downlink network.
We focus on linear precoding techniques, under the assumption that interference is treated as noise.
We assume all base stations share channel state information (CSI),
but a user's message is only routed to the base stations that participate in the user's coordination cluster.
The required backhaul bandwidth to share CSI between the base stations depends on the mobile speeds and update frequencies.
For typical applications, the requirement for sharing CSI is much less than that for sharing user data.
In particular, in \cite{samardzija09:backhaul_net_mimo}, it is shown that the requirement on backhaul bandwidth is about an order of magnitude less for pedestrian speeds.

The remainder of this paper is organized as follows.
The MIMO cellular network downlink model is presented in Section~\ref{sec:sys_mod}.
Section~\ref{sec:coop_cell_net} explains base station cooperation and the precoder optimization framework.
Section~\ref{sec:soft_if_null} describes the soft interference nulling algorithm and different clustering techniques.
Numerical results on the performance of different downlink precoding algorithms are presented in Section~\ref{sec:num_res}, under the settings of a line network and a cellular network.
Section~\ref{sec:conclu} concludes the paper.

\paragraph*{Notation}
In this paper, ($\mathds{R}_+$) $\mathds{R}$ is the set of (nonnegative) real numbers,
$\mathds{C}$ is the complex field, and $\mathds{1}$ denotes the two-element set $\{0,1\}$.
Dimensions of vectors/matrices are indicated by superscripts.
$\mathds{H}_+^N$ ($\mathds{H}_{++}^N$) is the set of $N \times N$ positive semidefinite (definite) Hermitian matrices.
$I_N$ is the $N \times N$ identity matrix;
$\mathbf{0}_{M\times N}$ is an $M\times N$ matrix with all zero entries.
$A^T$, $A^H$, $A^\dagger$ are the transpose, conjugate transpose, and pseudoinverse, respectively, of a matrix $A$.
$[A]_{i,j}$ is the matrix's $(i,j)$ entry, and $[a_{ij}]$ refers to the matrix comprising the entries $a_{ij}$.
The matrix $\diag(a)$ is diagonal, with its diagonal given by the vector $a$.
The operators $\E[\,\cdot\,]$, $\det$, $\tr$ denote, respectively, expectation, determinant, and trace.
For random variables, $x\sim\mathcal{CN}(\mu,Q)$, where $x,\mu\in \mathds{C}^N$, $Q\in\mathds{H}_+^N$,
means that $x$ is a circularly symmetric complex Gaussian random $N$-vector about mean $\mu$ with covariance matrix $Q$.

%%% ============================================================
\section{System Model}
\label{sec:sys_mod}

\begin{figure}
  \centering
  \psfrag{1}[][]{$1$}
  \psfrag{2}[][]{$2$}
  \psfrag{3}[][]{$3$}
  \psfrag{i}[][]{$i$}
  \psfrag{j}[][]{$j$}
  \psfrag{B}[][]{$B$}
  \psfrag{K}[][]{$K$}
  \psfrag{Pj}[][]{$P_j$}
  \psfrag{Mj}[][]{$M_j$}
  \psfrag{Ni}[][]{$N_i$}
  \psfrag{Hij}[][]{$H_{ij}$}
  \psfrag{K1}[][]{$\mathcal{K}_1$}
  \psfrag{K2}[][]{$\mathcal{K}_2$}
  \psfrag{K3}[][]{$\mathcal{K}_3$}
  \psfrag{B1}[l][l]{$\mathcal{B}_1$}
  \psfrag{B2}[l][l]{$\mathcal{B}_2$}
  \psfrag{B3}[l][l]{$\mathcal{B}_3$}
  \includegraphics{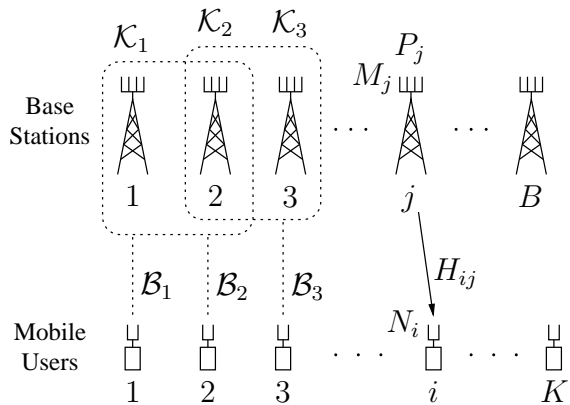}
  \caption{Downlink cooperative MIMO cellular network with limited coordination clusters.
  Base~$j$, with $M_j$ antennas, is under power constraint $P_j$; User~$i$, with $N_i$ antennas, has channel $H_{ij}$ from Base~$j$.
  In the figure, the coordination clusters are: $\mathcal{B}_1 = \mathcal{B}_2 = \{1,2\}$, $\mathcal{B}_3 = \{2,3\}$,
and the corresponding user sets are: $\mathcal{K}_1 = \{1,2\}$, $\mathcal{K}_2 = \{1,2,3\}$, $\mathcal{K}_3 = \{3\}$.
  }
  \label{fig:coop_cell_dl}
\end{figure}

Consider a MIMO downlink cellular network as depicted in Fig.~\ref{fig:coop_cell_dl}.
Suppose there are $B$ base stations and $K$ mobile users in the network.
Base~$j$ has $M_j$ transmit antennas, $j=1,\dotsc,B$; and User~$i$ has $N_i$ receiver antennas, $i=1,\dotsc K$.
We consider a narrow-band flat-fading channel model.
Wireless systems with wider bandwidth may be modeled as multiple narrow-band channels using modulation schemes such as orthogonal frequency-division multiplexing (OFDM), and most techniques discussed in this paper remain applicable.
The MIMO complex baseband channel gain from Base~$j$ to User~$i$ is denoted by the matrix $H_{ij} \in \mathds{C}^{N_i \times M_j}$.
Suppose $x_j \in \mathds{C}^{M_j}$ is the transmit signal at Base~$j$, and $y_i \in \mathds{C}^{N_i}$ is the receive signal at User~$i$,
then the discrete-time downlink channel is described by
\begin{align}
y_i &= \sum_{j=1}^B H_{ij} x_j + z_i,\quad i=1,\dotsc,K
\end{align}
where each $z_i \sim \mathcal{CN}(0,I_{N_i}) \in \mathds{C}^{N_i}$ is independent zero-mean circularly symmetric complex Gaussian (ZMCSCG) noise.

We consider a block-fading channel model:
the channel gains realize independently according to their distribution at the beginning of each fading block, and they remain unchanged within the duration of the fading block.
In this paper, we assume the channel states can be estimated accurately and conveyed in a timely manner to all base stations: i.e., the channels are known at all terminals.
Each Base~$j$ is under a transmit power constraint of $P_j$.
We consider a short-term power constraint: i.e., $\E[x_j^H x_j] \leq P_j$, $j=1,\dotsc,B$, where the expectation is over repeated channel uses within a fading block; power allocation across fading blocks is not considered.
We assume each fading block is sufficiently long so the transmitters may code at near channel capacity using random Gaussian codewords.

%%% ============================================================
\section{Cooperative Cellular Networks}
\label{sec:coop_cell_net}

\subsection{Base Station Coordination Clusters}
\label{sec:bs_coord_cluster}

Let us consider the scenario where a mobile user is served by a cluster of cooperative base stations.
Particularly, each User~$i$ specifies a coordination cluster $\mathcal{B}_i$,
where $\mathcal{B}_i \subseteq \{1,\dotsc,B\}$ is the set of base stations that participate in the cooperative transmission to User~$i$, $i=1,\dotsc,K$.
Different clustering techniques are discussed subsequently in Section~\ref{sec:clustering_alg}.
We assume all base stations share global CSI, i.e., all bases have knowledge of all channels $H_{ij}$.
Furthermore, the base stations in the coordination cluster $\mathcal{B}_i$ all know the message intended for User~$i$, and they may jointly encode the message in their transmission.
Conversely, the base stations unaffiliated with User~$i$'s coordination cluster $\{j \st j\notin \mathcal{B}_i\}$ do not have access to User~$i$'s message.
Let the total number of transmit antennas in coordination cluster $\mathcal{B}_i$ be denoted by
\begin{align}
\bar{M}_i \triangleq \sum_{j\in\mathcal{B}_i}M_j,\qquad i=1,\dotsc,K.
\end{align}
We use the notation $\mathcal{B}_i[n]$ to represent the $n$th element in the set $\mathcal{B}_i$ (sorted ascendingly),
and $\abs{\mathcal{B}_i}$ to denote the set's cardinality.
Note that the coordination clusters for different users may overlap, i.e., Base~$j$ may participate in the transmission to multiple users.
We denote the set of users served by Base~$j$ as
\begin{align}
\mathcal{K}_j &\triangleq \{i \st j \in \mathcal{B}_i\} \subseteq \{1,\dotsc,K\},\qquad j=1,\dotsc,B.
\end{align}
For example, the coordination clusters of Users~$1,2,3$, and their corresponding user sets, are illustrated in Fig.~\ref{fig:coop_cell_dl}.

\subsection{Linear Precoding}
\label{sec:linear_precoding}

In this paper, we consider linear precoding at each base station.
Specifically, let the transmit signal at Base~$j$ be given as follows:
\begin{align}
x_j &= \sum_{i\in\mathcal{K}_j} G_{ji}u_i
\end{align}
where $u_i\in\mathds{C}^{\bar{M}_i}$ represents the information signal from coordination cluster $\mathcal{B}_i$ to User~$i$,
and $G_{ji}\in\mathds{C}^{M_j\times\bar{M}_i}$ is the precoding matrix for User~$i$'s signal at Base~$j$.
Note that in the above formulation, we allow each User~$i$'s information signal to have multiple spatial streams
(up to the total number of transmit antennas $\bar{M}_i$ in its coordination cluster).
Without loss of generality, entries of a user's precoding matrix may be set to zero if the spatial streams are not all active.
Such representation is particularly useful under limited coordination clusters: since the clusters may overlap,
it may not be clear \emph{a priori} how many spatial streams are supported for each user, and how many antennas are used for beamforming vs.\ nulling out interference.
We assume each spatial stream consists of independent data signals, i.e., $\E[u_iu_i^H] = I_{\bar{M}_i}$,
and $\E[u_iu_k^H] = \mathbf{0}_{\bar{M}_i\times\bar{M}_k}$ for $i\neq k$.
It is convenient to specify the design of the precoding matrices $G_{ji}$ in terms of their corresponding covariance matrices
\begin{align}
Q_i &\triangleq G_iG_i^H \in \mathds{H}_+^{\bar{M}_i},&
G_i &\triangleq
\begin{bmatrix}
G_{\mathcal{B}_i[1]\,i}\\
G_{\mathcal{B}_i[2]\,i}\\
\vdots\\
G_{\mathcal{B}_i[\abs{\mathcal{B}_i}]\,i}
\end{bmatrix}
\in \mathds{C}^{\bar{M}_i\times \bar{M}_i}
\end{align}
where $i=1,\dotsc,K$,
and $\mathcal{B}_i[m]$ is the index of the $m$th base station in coordination cluster $\mathcal{B}_i$, with $m=1,\dotsc,\abs{\mathcal{B}_i}$.
To generate transmit signals with the specified covariances, we may set the precoding matrices $G_i$ to be
\begin{align}
\label{eq:Gi_ViDi_12}
G_i = V_iD_i^{\nicefrac{1}{2}}, \qquad i=1,\dotsc,K
\end{align}
where $D_i$ is diagonal, and $V_iD_iV_i^H = Q_i$ is the eigendecomposition of $Q_i$.

For each User~$i$, the desired signal is $\sum_{j\in\mathcal{B}_i} H_{ij}G_{ji}u_i$,
and the interference from the other users' signals is given by $\sum_{k=1,k\neq i}^K \sum_{j\in\mathcal{B}_k} H_{ij}G_{jk}u_k$.
Note in the above that each user suffers interference from all base stations: i.e., a frequency reuse factor of $1$ is assumed.
In this paper, we consider the design of linear precoders where interference is treated as noise
(i.e., interference pre-subtraction schemes such as dirty paper coding are not considered).
Let $R_i$ be the achievable rate for User~$i$.
When other users' signals are treated as noise, the following rate is achievable \cite{telatar99:cap_mimo_gaus, paulraj03:intro_st_wcom} using Gaussian signals
\begin{align}
\label{eq:Ri_logdet_HQH_1K}
R_i &= \log \frac{\det(I_{N_i} + \sum_{k=1}^K H_i C_kQ_kC_k^T H_i^H)}{\det(I_{N_i} + \sum_{k=1,k\neq i}^K H_i C_kQ_kC_k^T H_i^H)},
\qquad i = 1,\dotsc,K
\end{align}
where $H_i$ is the aggregate channel matrix from all base stations to User~$i$
\begin{align}
H_i &\triangleq
\begin{bmatrix}H_{i1} & H_{i2} & \dots & H_{iB}\end{bmatrix} \in \mathds{C}^{N_i \times \bar{M}},&
\bar{M} &\triangleq \sum_{j=1}^B M_j
\end{align}
and $C_k$ are constant matrices (each entry of $C_k$ is either a $0$ or $1$)
that represent the association between the base stations and the users in the coordination clusters
\begin{align}
\label{eq:Ci_def}
C_i &\triangleq
\begin{bmatrix}
E_{\mathcal{B}_i[1]} & E_{\mathcal{B}_i[2]} & \dots & E_{\mathcal{B}_i[\abs{\mathcal{B}_i}]}
\end{bmatrix}
\in \mathds{1}^{\bar{M} \times \bar{M}_i},
\qquad i = 1,\dotsc,K
\end{align}
where
\begin{align}
\label{eq:Ej_def}
E_j \triangleq
\begin{bmatrix}
\mathbf{0}_{M_1\times M_j}\\
\vdots\\
I_{M_j}\\
\vdots\\
\mathbf{0}_{M_B\times M_j}\\
\end{bmatrix}
\in \mathds{1}^{\bar{M} \times M_j},
\qquad j = 1,\dotsc,B.
\end{align}
The $\bar{M}\times \bar{M}$ matrix $C_iQ_iC_i^T$ represents the full covariance matrix of User~$i$'s signal with respect to the transmit antennas of all base stations in the network.
Effectively, the matrix $C_i$ stipulates that the non-participating base stations have precoding weights of zero for User~$i$'s signal.
In terms of the matrices $C_i$ and $E_j$,
the transmit power constraint for Base~$j$ can be written as
\begin{align}
\sum_{i=1}^K \tr (E_j^TC_iQ_iC_i^TE_j) &\leq P_j,\qquad j = 1,\dotsc,B.
\end{align}
To illustrate the construction of the association matrices $C_i$, consider, for example, a cellular network with $B=5$ base stations, with each base station having a single antenna.
Suppose the coordination cluster for User~$i$ is $\mathcal{B}_i = \{1,3,4\}$.
Then User~$i$'s signal is characterized by the covariance matrix $Q_i\in\mathds{H}_+^3$,
which describes the joint signal from base stations $1,3,4$.
The association matrix $C_i$ and the full covariance matrix $C_iQ_iC_i^T$, respectively, are
\begin{align}
C_i &= \begin{bmatrix}
1 & 0 & 0\\
0 & 0 & 0\\
0 & 1 & 0\\
0 & 0 & 1\\
0 & 0 & 0
\end{bmatrix},&
C_iQ_iC_i^T &= \begin{bmatrix}
[Q_i]_{1,1} & 0 & [Q_i]_{1,2} & [Q_i]_{1,3} & 0\\
0 & 0 & 0 & 0 & 0\\
[Q_i]_{2,1} & 0 & [Q_i]_{2,2} & [Q_i]_{2,3} & 0\\
[Q_i]_{3,1} & 0 & [Q_i]_{3,2} & [Q_i]_{3,3} & 0\\
0 & 0 & 0 & 0 & 0
\end{bmatrix}.
\end{align}
Given the aggregate channel matrices, association matrices, and covariance matrices
($H_k$, $C_k$, $Q_k$, for $k=1,\ldots,K$),
the MIMO rate $R_i$ in (\ref{eq:Ri_logdet_HQH_1K}) can be achieved
through singular value decomposition (SVD) \cite{telatar99:cap_mimo_gaus},
or minimum mean square error (MMSE) detection with
successive interference cancellation (SIC) \cite{foschini98:wcomm_mult_ant}.
The SVD and MMSE-SIC strategies are capacity-achieving (when interference is treated as noise) for arbitrary numbers of transmit/receiver antennas, and number of spatial streams being active.

\subsection{Optimal Precoder}

Let the user rates in (\ref{eq:Ri_logdet_HQH_1K}) be denoted by the vector $R \triangleq [R_1 \,\dots\, R_K]^T \in\mathds{R}_+^K$.
We are interested in maximizing a concave utility function of the user rates.
Let $U : \mathds{R}^K_+ \rightarrow \mathds{R}$ denote the utility function, where $U$ is concave on $\mathds{R}_+^K$.
Concave utility functions can be used to model a wide class of resource allocation preferences among the users.
For example, we may consider the weighted sum of rates
\begin{align}
\label{eq:Uw_R}
U_w(R) &= \sum_{i=1}^K w_iR_i
\end{align}
where $w_i \geq 0$, $i=1,\dotsc,K$.
When the weights are all equal to unity in (\ref{eq:Uw_R}), i.e., $w_1 = \dotsb = w_K = 1$, the utility function is referred to as the sum rate.

To find the optimal linear precoders,
the design variables are the covariance matrices $Q_1,Q_2,\dotsc,Q_K$; the precoding matrices $G_i$ are recovered from the covariance matrices according to (\ref{eq:Gi_ViDi_12}).
The optimization problem can be formulated as
\begin{align}
\label{eq:max_U_R}
\text{maximize} \quad & U(R)\\
\label{eq:max_U_R_Qi}
\text{over} \quad & R\in\mathds{R}_+^K,\; Q_i \in \mathds{H}_+^{\bar{M}_i}\\
\text{subject to} \quad
\label{eq:max_Ri_log_det}
& R_i \leq \log \frac{\det(I_{N_i} + \sum_{k=1}^K H_i C_kQ_kC_k^T H_i^H)}{\det(I_{N_i} + \sum_{k=1,k\neq i}^K H_i C_kQ_kC_k^T H_i^H)}\\
\label{eq:max_U_R_Pj}
& \sum_{i=1}^K \tr (E_j^TC_iQ_iC_i^TE_j) \leq P_j
\end{align}
where $i=1,\dotsc,K$; $j=1,\dotsc,B$; and $R\triangleq[R_1 \,\dots\, R_K]^T$.
Note that within the scope of the optimization problem (\ref{eq:max_U_R})--(\ref{eq:max_U_R_Pj}), $R, Q_i$ as specified in (\ref{eq:max_U_R_Qi}) are simply placeholder variables;
they may take on any values in their respective domains as long as all constraints are satisfied.
However, the maximization in (\ref{eq:max_U_R})--(\ref{eq:max_U_R_Pj}) is not a convex optimization problem;
in general, it is difficult to compute the optimal $R$ and $Q_i$ efficiently.

%%% ============================================================
\section{Soft Interference Nulling}
\label{sec:soft_if_null}

\subsection{Precoder Optimization}
\label{sec:sin_precod_opt}

In this section, we propose a linear precoding technique called soft interference nulling (SIN),
which has good performance in the sense that SIN precoding performs better than or equal to any linear precoding scheme that aims to completely eliminate interference.
The SIN formulation is based on convexifying the optimization problem (\ref{eq:max_U_R})--(\ref{eq:max_U_R_Pj}) about some given operating point: $\bar{Q}_i \in \mathds{H}_+^{\bar{M}_i}$, $i=1,\dotsc,K$.
Specifically, for the inequality constraints in (\ref{eq:max_Ri_log_det}), we consider the first-order Taylor series expansion of the log-determinant function about $\bar{Q}_i$, $i=1,\dotsc,K$
\begin{align}
\label{eq:Ri_eq_log_tr_Q}
R_i
&= \log \det\Bigl(I_{N_i} + \sum_{k=1}^K H_i C_kQ_kC_k^T H_i^H\Bigr) - \log \det\Bigl(I_{N_i} + \sum_{k=1,k\neq i}^K H_i C_kQ_kC_k^T H_i^H\Bigr)\\
\label{eq:Ri_gts_log_tr_Qb}
\begin{split}
&\gtrsim \log \det\Bigl(I_{N_i} + \sum_{k=1}^K H_i C_kQ_kC_k^T H_i^H\Bigr) - \sum_{k=1,k\neq i}^K \tr (Y_i^{-1}H_i C_kQ_kC_k^T H_i^H)\\
&\qquad - \log \det Y_i
+ \sum_{k=1,k\neq i}^K \tr (Y_i^{-1}H_i C_k\bar{Q}_kC_k^T H_i^H)
\end{split}\\
\label{eq:def_tilde_Ri}
&\triangleq \tilde{R}_i
\end{align}
where $Y_i$ represents the estimated covariance of the interference plus noise
\begin{align}
\label{eq:Y_i}
Y_i \triangleq I_{N_i} + \sum_{k=1,k\neq i}^K H_i C_k\bar{Q}_kC_k^T H_i \quad \in \mathds{H}_{++}^{N_i}.
\end{align}
Note that when $Q_i$'s are near $\bar{Q}_i$'s, $R_i$'s are well approximated by $\tilde{R}_i$'s: equality holds when $Q_i=\bar{Q}_i$, $i=1,\dotsc,K$.
Moreover, the inequality in (\ref{eq:Ri_gts_log_tr_Qb}) is due to $\tr X$ being a global over-estimator of $\log\det(I_n+X)$,
which follows from the concavity of the log-determinant function \cite{boyd04:convex_opt}.

We consider the precoding matrices that correspond to the solution of following optimization problem
\begin{align}
\label{eq:max_URt_Qb}
\text{maximize} \quad & U(\tilde{R})\\
\text{over} \quad & \tilde{R}\in\mathds{R}_+^K,\; Q_i \in \mathds{H}_+^{\bar{M}_i}\\
\begin{split}
\text{subject to} \quad
\label{eq:max_URt_log_det_tr}
& \tilde{R}_i \leq \log \det\Bigl(I_{N_i} + \sum_{k=1}^K H_i C_kQ_kC_k^T H_i^H\Bigr) - \sum_{k=1,k\neq i}^K \tr (Y_i^{-1} H_i C_kQ_kC_k^T H_i^H)\\
&\qquad - \log \det Y_i + \sum_{k=1,k\neq i}^K \tr (Y_i^{-1}H_i C_k\bar{Q}_kC_k^T H_i^H)
\end{split}\\
\label{eq:max_URt_Qb_trPj}
& \sum_{i=1}^K \tr (E_j^TC_iQ_iC_i^TE_j) \leq P_j
\end{align}
where $i=1,\dotsc,K$; $j=1,\dotsc,B$;
the constant matrices $C_i$, $E_j$ are as given in (\ref{eq:Ci_def}), (\ref{eq:Ej_def});
and $Y_i$ is as defined in (\ref{eq:Y_i}).
Note that the maximization (\ref{eq:max_URt_Qb})--(\ref{eq:max_URt_Qb_trPj}) above is a convex optimization problem,
and its solution can be efficiently computed using standard convex optimization numerical techniques,
e.g., by the interior-point method \cite{renegar01:math_ipm_cvxopt, boyd04:convex_opt}.
The number of active spatial streams for User~$i$,
which corresponds to the rank of its input covariance matrix $Q_i$, is determined by the optimization framework.
In the optimization formulation above,
though there is no explicit rank constraint on $Q_i$,
we do not expect the effective rank of its solution $Q_i^*$ be greater than $N_i$, i.e., a user cannot receive more spatial streams than it has number of receive antennas.

In the following, we consider an iterative algorithm, as listed below in Algorithm~\ref{alg:iter_sin}.
First, we initialize $\bar{Q}_i$ to be zero matrices, $i = 1,\dotsc,K$, and solve (\ref{eq:max_URt_Qb})--(\ref{eq:max_URt_Qb_trPj}).
In the initial iteration, (\ref{eq:Ri_gts_log_tr_Qb}) is a good approximation for small interference levels.
In the next iteration, each $\bar{Q}_i$ takes on the value of $Q_i^*$ (Algorithm~\ref{alg:iter_sin}, Line~\ref{alg:update_R_Q}), the solution of the optimization problem (\ref{eq:max_URt_Qb})--(\ref{eq:max_URt_Qb_trPj}).
As we apply the first-order Taylor series expansion about the updated $\bar{Q}_i$ in the new iteration,
accordingly, the formulation in (\ref{eq:Ri_gts_log_tr_Qb}) becomes a good approximation about the updated interference levels.
This procedure is repeated to iteratively refine the estimated covariance $Y_i$ of the realized interference plus noise in (\ref{eq:Y_i}).
Subsequently, we show that the iterations are monotonically nondecreasing, and we terminate the algorithm if the utility improvement is less than a given tolerance $\varepsilon > 0$.
Therefore, when the algorithm terminates, the achievable rates $R_i$ (\ref{eq:Ri_eq_log_tr_Q}) are locally well-approximated (to the first order) by the convexified rate expressions $\tilde{R}_i$ in (\ref{eq:def_tilde_Ri}) about the realized interference levels.
At the conclusion of the algorithm, the realized interference-plus-noise covariance is represented by $Y_i$;
there is no stipulation on the interference levels being small.
The approximation formulation in (\ref{eq:Ri_gts_log_tr_Qb}) is valid for arbitrary numbers of transmit and receive antennas, and arbitrary levels of interference in the network.

\begin{algorithm}
\caption{Iterative Precoder Optimization}
\label{alg:iter_sin}
\begin{algorithmic}[1]
\State Initialize:
$\bar{R} \leftarrow \mathbf{0}_{K\times1}$,\; $\bar{Q}_i \leftarrow  \mathbf{0}_{\bar{M}_i\times\bar{M}_i}$,\quad $i=1,\dotsc,K$
\Loop
    \State Compute $\tilde{R}^*,Q_i^*,\,i=1,\dotsc,K$, by solving (\ref{eq:max_URt_Qb})--(\ref{eq:max_URt_Qb_trPj})
    \If{$U(\tilde{R}^*)-U(\bar{R}) < \varepsilon$}
        \State \textbf{break}
    \Else
        \State $\bar{R} \leftarrow \tilde{R}^*$,\; $\bar{Q}_i \leftarrow Q_i^*$,\quad $i=1,\dotsc,K$
\label{alg:update_R_Q}
    \EndIf
\EndLoop
\end{algorithmic}
\end{algorithm}

We now show that the SIN precoding algorithm performs at least as well as any linear precoding scheme that completely eliminates all interference.
Recall from Section~\ref{sec:linear_precoding}, a linear precoding scheme is \emph{interference-free} if
\begin{align}
\label{eq:intf_free_cond}
\sum_{k=1,k\neq i}^K \sum_{j\in\mathcal{B}_k} H_{ij}G_{jk}u_k = \mathbf{0}_{N_i\times1}, \quad i=1,\dotsc,K.
\end{align}
For example, zero-forcing beamforming achieves the above interference-free condition.
\begin{prop}
\label{prop:SIN_intf_free}
The SIN precoding scheme given in Algorithm~\ref{alg:iter_sin} performs better than or equal to any linear interference-free precoding scheme.
\end{prop}
\begin{IEEEproof}
Suppose a linear interference-free precoding scheme has precoding matrices: $\hat{Q}_i$, $i=1,\dotsc,K$.
Then User~$i$ achieves the rate
\begin{align}
\label{eq:R_hat_Q_hat}
\hat{R}_i &= \log \det(I_{N_i} + H_i C_i\hat{Q_i}C_i^T H_i^H)
\end{align}
where (\ref{eq:R_hat_Q_hat}) follows from substituting the interference-free condition (\ref{eq:intf_free_cond}) in (\ref{eq:Ri_logdet_HQH_1K}).
Note that in the initial iteration in Algorithm~\ref{alg:iter_sin}, $\bar{Q}_i = \mathbf{0}_{\bar{M}_i\times\bar{M}_i}$,
and hence $Y_i = \mathbf{0}_{N_i\times N_i}$, $i=1,\dotsc,K$.
Substituting (\ref{eq:intf_free_cond}), (\ref{eq:R_hat_Q_hat}) in (\ref{eq:max_URt_log_det_tr}),
it is seen that $\hat{R}$, $\hat{Q}_i$ are feasible in the SIN optimization problem (\ref{eq:max_URt_Qb})--(\ref{eq:max_URt_Qb_trPj}).
Naturally, the solution of the optimization problem is better than or equal to any feasible solution.
Furthermore, since $\tilde{R}_i$ is a global under-estimator of $R_i$, the objective function can only improve (or stay the same) after each iteration.
As the user rates are bounded, the algorithm converges to a local optimum.
Consequently, when the algorithm terminates, the SIN precoder performs at least as well as the interference-free precoder.
\end{IEEEproof}
An interference-free precoding scheme can be interpreted as one that imposes an infinite penalty on interference, whereas SIN relaxes such restriction and allows the possibility of nonzero interference.
Moreover, SIN precoding is well-defined even when the number of transmit antennas is less than the total number of receive antennas.
A separate user selection step is not necessary: under SIN precoding, the set of active users (i.e., those with nonzero rates) is determined by Algorithm~\ref{alg:iter_sin}.

%%% ============================================================
\subsection{Clustering Algorithms}
\label{sec:clustering_alg}

In Section~\ref{sec:bs_coord_cluster}, we assume the coordination cluster $\mathcal{B}_i$ for each User~$i$ is given.
As exhaustive search for the best clustering combinations has complexity too high even for small network sizes,
in this section, we consider two simple clustering techniques.
The users choose their coordination clusters based on the long-term channel conditions (i.e., the shadowing realizations), but they cannot adapt the coordination clusters based on the fast Rayleigh fading realizations.
In each algorithm, we assume the coordination cluster size $\abs{\mathcal{B}_i}>1$ for each User~$i$ is given.
\paragraph{Nearest Bases Clustering}
In this simple clustering scheme, each User~$i$ chooses the nearest (in terms of signal strength) $\abs{\mathcal{B}_i}$ bases to comprise its coordination cluster.
\paragraph{Nearest Interferers Clustering}
In the nearest interferers clustering scheme, each user attempts to reduce interference to its $\abs{\mathcal{B}_i}-1$ closest neighbors.
First, each User~$i$ chooses the nearest base (in terms of signal strength) as its home base station.
Let $B_i$ denote the home base station of User~$i$.
Next, let $\mathcal{I}_i$ denote the set of neighbors of User~$i$ who suffer the most interference from $B_i$.
User~$i$ then forms its coordination cluster as $B_i \,\bigcup\, \{B_k \st k\in\mathcal{I}_i\}$.

%%% ============================================================
\section{Numerical Results}
\label{sec:num_res}

\subsection{Line Network}
\label{sec:line_net}

\subsubsection{Network Geometry}
\begin{figure}
  \centering
  \psfrag{1}[][]{$1$}
  \psfrag{2}[][]{$2$}
  \psfrag{3}[][]{$3$}
  \psfrag{B}[][]{$B$}
  \psfrag{dx}[][]{$d_x$}
  \psfrag{dy}[][]{$d_y$}
  \includegraphics{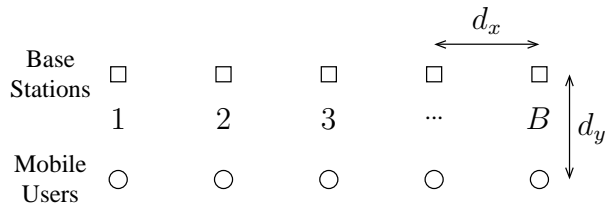}
  \caption{Line network model.}
  \label{fig:cell_line_net}
\end{figure}

We first consider a simple line network model to gain intuition on the behavior of different downlink precoding algorithms.
Suppose there are $B$ base stations in the network, and they are positioned along a line with distance $d_x$ apart.
Each base station serves one mobile user, and we assume each user is located at a distance $d_y$ away from its base station, as illustrated in Fig.~\ref{fig:cell_line_net}.
The total number of users in the network thus is $K = B$.
To minimize the boundary effects, we consider a line network with wraparound where the distance $d_{ij}$ between User~$i$ and Base~$j$ is given by
\begin{align}
d_{ij} &\triangleq \sqrt{d_y^2+\bigl(d_x d(i,j)\bigr)^2}, \qquad i,j=1,\dotsc,B
\end{align}
where
\begin{align}
d(i,j) &\triangleq d \in \{0,1,\dotsc,B-1\} \;\st\; d \equiv i-j \pmod{B}.
\end{align}

The radio signal propagation from Base~$j$ to User~$i$ is modeled as independent Rayleigh fading with a distance-based power attenuation of $d_{ij}^{-\eta}$, where $\eta$ is the path loss exponent:
i.e., each channel entry is independent and identically distributed (i.i.d.) as $\mathcal{CN}(0,d_{ij}^{-\eta})$.
We assume the line network has geometry: $d_x=d_y=1$, and $\eta=4$.
In the line network, we consider the case where each base station and each user has a single antenna,
i.e., $M_j = N_i = 1$, for all $i=1,\dotsc,K$, and $j=1,\dotsc,B$.
For such single-input single-output (SISO) channels, the channel gain is a scalar, and we will use the notation
\begin{align}
h_{ij} \triangleq H_{ij} \in\mathds{C},\qquad i=1,\dotsc,K,\quad j=1,\dotsc,B.
\end{align}

\subsubsection{Dirty Paper Coding}

For collaborative cellular networks, a performance upper bound is obtained when all base stations cooperate perfectly.
Suppose each base station knows the messages of all users, and we allow joint encoding at the base stations.
Then this cooperative cellular system may be modeled as a broadcast channel (BC) with $K$ single-antenna receivers, and an $B$-antenna transmitter under per-antenna power constraints (PAPC).
For a Gaussian MIMO BC, its capacity region \cite{weingarten06:cap_mimo_bc} is achieved by the dirty paper coding (DPC) scheme \cite{costa83:writing_dirty_paper}.
In DPC, the messages for the users are encoded in a given order, and the interference from the previously encoded users is pre-subtracted at the transmitter for the subsequently encoded users.
In \cite{yu07:tx_opt_ma_dl_papc}, it is shown that the sum rate of a MIMO BC under PAPC can be computed by solving the following convex minimax optimization problem:
\begin{align}
\min_{q}\; \max_{s} \quad &\log \det \Bigl(\sum_{i=1}^K s_i\tilde{h}_i\tilde{h}_i^H + \diag(q)\Bigr) - \sum_{i=1}^B \log q_i\\
\text{over} \quad & q\in \mathds{R}_+^B,\; s\in \mathds{R}_+^K\\
\text{subject to} \quad
& \sum_{i=1}^K s_i \leq \sum_{j=1}^B P_j\\
& \sum_{i=1}^B q_i P_j \leq \sum_{j=1}^B P_j
\end{align}
where
\begin{align}
\label{eq:htil_q_s}
\tilde{h}_i &\triangleq [h_{i1} \dots h_{iB}]^T \in\mathds{C}^B, &
q &\triangleq [q_1 \dots q_B]^T \in\mathds{R}^B_+, &
s &\triangleq [s_1 \dots s_K]^T \in\mathds{R}^K_+.
\end{align}

\subsubsection{Full-Network Zero-Forcing}

In the case where \emph{all} base stations participate in the joint encoding of the user messages,
a simple linear transmit precoding technique is zero-forcing (ZF) beamforming \cite{caire03:ach_guas_mimo_bc}.
Unlike DPC, no interference pre-subtraction is performed at the transmitter.
Instead, the transmitter sets the precoder to be the pseudoinverse of the channel matrix, such that interference is zeroed out at each mobile user.
Zero-forcing in MIMO BC subject to PAPC is considered in \cite{boccardi06:zf_mimo_bc_papc, karakayali07:opt_zf_papc, wiesel08:zf_pre_gen_inv}.
In particular, the ZF sum rate under PAPC can be found by solving the following convex optimization problem:
\begin{align}
\text{maximize}\quad & \sum_{i=1}^K \log(1+\gamma_i)\\
\text{over}\quad & \gamma \in \mathds{R}_+^K\\
\label{eq:max_ZF_W_P}
\text{subject to}\quad &
\lvert W \rvert^2 \gamma \leq [P_1 \,\dots\, P_B]^T
\end{align}
where $\gamma \triangleq [\gamma_1 \dots \gamma_K]^T \in\mathds{R}^K_+$; $W = [h_{ij}]^{\dagger}$;
the constraint in (\ref{eq:max_ZF_W_P}) represents component-wise inequality; and
$\lvert W \rvert^2$ denotes the component-wise squared magnitude of the entries of $W$, i.e., $\lvert W \rvert^2 \triangleq [\lvert w_{ij}\rvert^2]$.
In general, ZF is suboptimal. However, at high SNR with multiuser diversity, ZF performs asymptotically close to the optimal DPC scheme
\cite{sharif05:mimo_bc_part_si, yoo06:opt_ma_bc_zf_bf, sharif07:ts_dpc_bf_mimo_bc_users}.

\subsubsection{Achievable Rates and Capacity Bounds}

We consider the sum rate of the downlink channel, normalized by the number of base stations. All the base stations are under the same transmit power constraints: $P_1 = \dotsb = P_B = P$,
and we also refer to $P$ as the SNR of the system.
In the numerical experiments, $100$ sets of random channel realizations are generated.
For each set of channel realizations, the non-cooperative and the cooperative normalized sum rates are calculated.
Then the rates are averaged over the random channel realizations.
The convex optimization problems are solved using the software package SDPT3 \cite{tutuncu03:sdpt3}.

\begin{figure}
  \centering
  \includegraphics*[width=10cm]{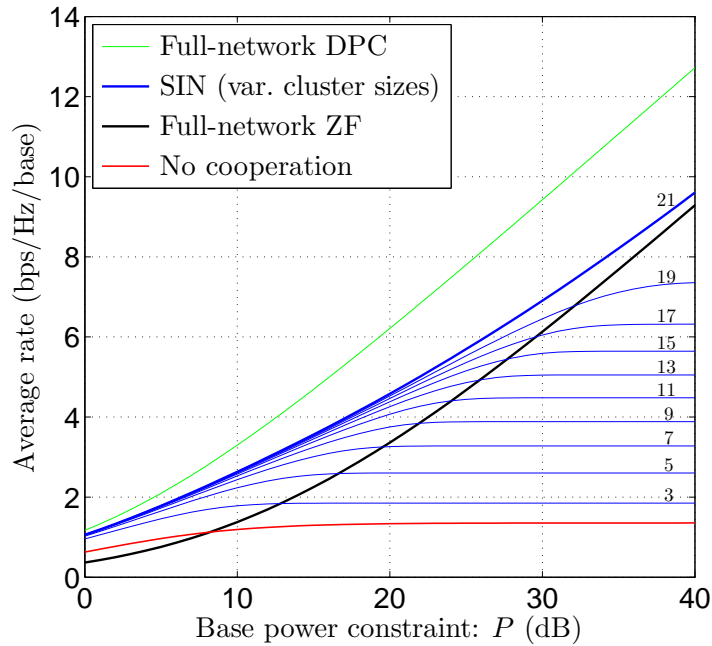}
  \caption{Soft interference nulling (SIN) with different coordination cluster sizes (labeled next to their corresponding curves) and full-network zero-forcing (ZF) rates in a line network ($B=K=21$, $d_x=d_y=1$),
  as compared to full-network DPC and non-cooperative full-power transmission.}
  \label{fig:sin_B21_M1_N1_R100}
\end{figure}

The line network achievable rates and capacity bounds are shown in Fig.~\ref{fig:sin_B21_M1_N1_R100} as a function of the SNR $P$, with $B=K=21$.
In the line network, for simplicity, all users are assumed active and the SIN rates are computed with a single iteration of Algorithm~\ref{alg:iter_sin}.
The non-cooperative baseline refers to the case where each base station transmits at full power.
Without base station cooperation, the average user rate saturates at approximately $1.6\,\text{bps}/\text{Hz}$.
On the other hand, interference can be overcome by allowing the base stations to cooperate, as shown by the DPC rates when the cooperative system is modeled as a BC.\@
ZF beamforming is able to achieve increasing system throughput as the SNR improves.
There is a gap between between the ZF rate and the DPC rate, but they both exhibit similar scaling trends as $P$ increases.
In particular, the ZF beamforming technique allows the network to overcome its interference-limited performance bottleneck, thus demonstrating the value of cooperative cellular networks.
However, ZF beamforming requires all base stations in the network to cooperate.

\subsubsection{Soft Interference Nulling Precoding}
\label{sec:line_net_sin}

For the SIN precoding scheme, different coordination cluster sizes are considered, and the cluster sizes are labeled next to their corresponding curves on the plot.
For the line network topology, the two clustering algorithms discussed in Section~\ref{sec:clustering_alg} are equivalent:
i.e., each user chooses the closest $\abs{\mathcal{B}_i}$ base stations to participate in its coordination cluster.
For example, when the coordination cluster size is $3$, User~$1$ is served by Bases~$\{21,1,2\}$, User~$2$ by Bases~$\{1,2,3\}$, User~$3$ by Bases~$\{2,3,4\}$, and so on.
A coordination cluster size of $21$ represents full network coordination.

Note that the full-network ZF beamforming scheme satisfies the interference-free condition (\ref{eq:intf_free_cond});
therefore, under full network coordination, by Proposition~\ref{prop:SIN_intf_free}, SIN precoding outperforms ZF\@.
In particular, at low SNRs, ZF suffers from noise amplification while SIN does not.
Moreover, at moderate SNRs, SIN with limited coordination cluster sizes is able to outperform ZF with full network coordination.
For example, at the SNR $P = 18\,\dB$,
SIN with a coordination cluster size of $7$ outperforms ZF with full network coordination.
As the SNR increases, however, it is observed that SIN under partial network coordination once again becomes interference-limited.

\subsection{Cellular Network}
\label{sec:cell_net}

\subsubsection{Network Geometry}
\begin{figure}
  \centering
  \includegraphics[scale=1.5]{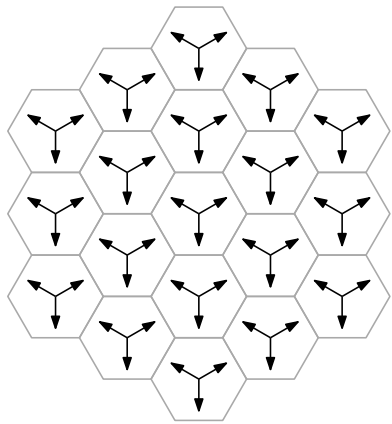}
  \caption{Hexagonal three-sectored cellular wireless network. There are $19$ cells in the network, with wraparound at the edges. Each cell has three sectors.
  Each sector corresponds to a base station, and each base station serves one mobile.
  Each arrow represents the boresight direction of a base station's antenna beam.}
  \label{fig:hex_cell}
\end{figure}

We next consider a cellular network, as shown in Fig.~\ref{fig:hex_cell}, which consists of two rings of hexagonal cells, with wraparound at the boundary.
Each cell has three sectors; thus there are $19$ cells, or $57$ sectors, in the network.
Each sector has one transmit antenna and serves one user, and each user has a single receive antenna.
Users are randomly populated in the network.
The distance between any two closest cell centers is $0.5\,\mathrm{km}$.
Average channel SNR is determined by propagation path-loss (with a path-loss exponent of $3.76$) and shadow fading
(with $8$-$\dB$ standard deviation, $50$-$\mathrm{m}$ correlation distance, $50\%$ correlation across sites).
The transmit antenna at each sector has a parabolic beam pattern.
A user is associated with the sector to which it has the highest average SNR (up to the maximum of one user per sector).
The users are indexed such that Base~$i$ wishes to transmit to User~$i$.
In composition with the path-loss and shadowing, each channel also experiences i.i.d. fast Rayleigh fading.
Each sector is under a transmit power constraint that corresponds to a cell-edge SNR of $20\,\dB$.

\subsubsection{Myopic Zero-Forcing}

Zero-forcing requires all base stations to cooperate in order to ensure an interference-free signal at each of the users.
However, in a wide-area cellular network, it is unrealistic to demand full-network coordination.
When only clusters of base stations cooperate, interference cannot be completely eliminated.
For comparison, in this section,
a simple myopic scheme is considered.
First, in the case when a base station belongs to multiple clusters, it divides its transmit power equally among the clusters.
Then, within each cluster, the ZF precoder is used.
Finally, the inter-cluster interference is treated as noise when the achievable rates are computed.
Myopic ZF requires less backhaul bandwidth in exchanging CSI;
however, \cite{samardzija09:backhaul_net_mimo} shows that the backhaul overhead is dominated by data sharing,
for which the same bandwidth requirements apply for myopic ZF and SIN\@.

\subsubsection{Performance Comparison}

\begin{figure}
  \centering
  \includegraphics*[width=10cm]{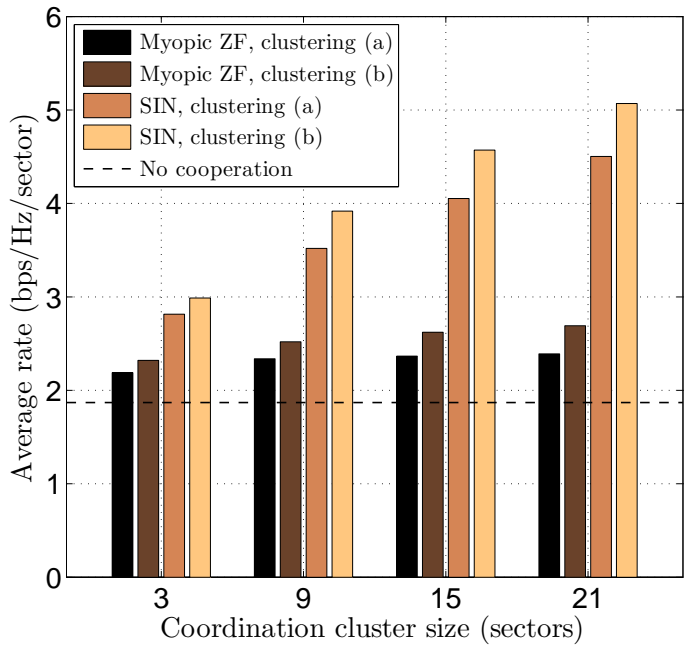}
  \caption{Average rates under different transmission strategies for various coordination cluster sizes.
  Clustering~(a) refers to nearest bases clustering, and (b) refers to nearest interferers clustering.}
  \label{fig:nRings2_nS10_nR10_avg_rate}
\end{figure}

For the numerical experiments, $10$ instances of shadow fading realizations are generated.
For each shadow fading realization, $10$ Raleigh fading instances are generated (i.e., a total of $100$ sets of channel realizations).
In the cellular network, we consider a simple user selection mechanism.
In calculating the myopic ZF rates, the users that achieve the lowest $10\%$ non-cooperative rates are allowed to be in outage.
To compute the SIN precoding matrices, iterations of Algorithm~\ref{alg:iter_sin} are carried out until a tolerance of $\varepsilon=0.01$ is achieved (i.e., the set of active users is determined by the algorithm).
The average rates in cellular network under different clustering algorithms for various coordination cluster sizes are shown in Fig.~\ref{fig:nRings2_nS10_nR10_avg_rate}.
In the figure, clustering~(a), (b) refer to nearest bases clustering, and nearest interferers clustering, respectively, as described in Section~\ref{sec:clustering_alg}.
It is observed that nearest interferers clustering outperforms nearest bases clustering, which suggests cooperation is more useful in mitigating interference than boosting the users' signal strength.
Unfortunately, the myopic ZF scheme does not satisfy the interference-free condition in (\ref{eq:intf_free_cond}), and Proposition~\ref{prop:SIN_intf_free} does not apply in the comparison of its performance with SIN precoding.
However, the simulation results show that
for a given coordination cluster size, SIN precoding achieves considerably higher throughput compared with myopic ZF\@.
Furthermore, SIN can more effectively take advantage of larger coordination clusters:
the SIN rates improve when the coordination clusters become larger; whereas the improvement in the myopic-ZF rates is only marginal.
Intuitively, SIN precoding is able to achieve better performance over myopic ZF because
i) SIN optimizes the transmit power allocation among the coordination clusters,
and ii) SIN aims to achieve the optimal balance between beamforming and interference nulling in using the multiple antennas in each cluster.

%%% ============================================================
\section{Conclusions}
\label{sec:conclu}

In this paper, we consider maximizing a concave utility function of the user rates in a cooperative MIMO cellular downlink network.
Without base station cooperation, a cellular network is interference-limited.
Cooperation among base stations allows the joint encoding of user signals, which can overcome the interference limitation.
However, the capacity-achieving dirty paper coding (DPC) scheme has high complexity.
When all base stations in the network cooperate, zero-forcing (ZF) beamforming offers good performance relative to DPC\@.
We investigate the case where each user may specify only a subset of the base stations to form a coordination cluster, and different coordination clusters may overlap in the network.
The base stations share CSI, but a user's message is only routed to base stations in the coordination cluster associated with the user.

We focus on linear precoding techniques for low-complexity implementation.
In general, the optimal linear precoding is a nonconvex problem which is difficult to compute efficiently.
We propose an approximation formulation called soft interference nulling (SIN), which performs as well as or better than ZF beamforming.
The SIN precoding matrices are computed by solving a sequence of convex optimization problems.
In a line network,
it is shown that SIN precoding under partial network coordination can outperform ZF under full network coordination at moderate SNRs.
In a hexagonal three-sectored cellular network, SIN precoding achieves considerably higher throughput compared to myopic ZF with the same coordination cluster size.
Moreover, the SIN rates improve with increasing coordination cluster sizes, while the myopic-ZF rates do so only marginally.

%%% ============================================================
\section*{Acknowledgment}
The authors would like to thank Dennis R. Morgan for helpful discussions.

%%% ============================================================
\bibliographystyle{IEEEtran}
\bibliography{IEEEabrv,ref}

\end{document}